\documentclass[twocolumn,showpacs,preprintnumbers,amsmath,amssymb,superscriptaddress]{revtex4}
\usepackage{graphicx}
\begin{document}

\title{Magnetic anisotropy of single 3$d$ spins on CuN surface }

\author{A. B. Shick}
\author{F. M\'aca}
\affiliation{Institute of Physics ASCR, Na Slovance 2, Prague 8,
Czech Republic}
\author{A. I. Lichtenstein}
\affiliation{University of Hamburg, Jungiusstrasse 9, 20355 Hamburg,
Germany}

\begin{abstract}
First-principles calculations of the magnetic anisotropy energy for
Mn- and Fe-atoms on CuN/Cu(001) surface are performed making use of
the torque method. The easy magnetization direction is found to be
different for Mn and Fe atoms in accord with the experiment. It is
shown the magnetic anisotropy has a single-ion character and mainly
originates from the local magnetic moment of Mn- and Fe-atoms. The
uniaxial magnetic anisotropy constants are calculated in reasonable
agreement with the experiment.
\end{abstract}
\date{\today}
\maketitle

Recent  scanning tunneling miscropscopy (STM) measurements of the
spin-excitation energies in a magnetic field \cite{Hirjibehedin} for
individual Fe and Mn atoms on CuN/Cu(001)-c(2x2) substrate report
large values of the axial and transverse magnetic anisotropy
energies (MAE) for a single magnetic atom. The STM experiments are
complemented by density-functional theoretical calculations. These
calculations reveal that the magnetic atoms become incorporated into
a covalent CuN matrix, so that their electronic and magnetic
character differs from the gas-phase transition metal atoms.

These STM experiments \cite{Hirjibehedin} along with previously
reported XMCD measurements \cite{gambardella03} for a single Co atom
and small Co clusters on the Pt(111) surface show that just a few
atom size nanostructures can maintain a stable magnetic orientation
at low temperature due to the large magnetic anisotropy energy
(MAE). What makes these atomic-scale magnetic structures
technologically relevant is their large MAE which provides the means
of reducing the size of the magnetic bits above the
superparamagnetic limit, i.e. the ratio of the MAE to the thermal
energy $k_BT$.
Understanding of the atomic-scale MAE in nanomagnets is essential in
the determination of the minimum feasible magnetic memory bit size,
and can assist in further increase of the magnetic recording
density.

In the work reported here we make use of  {\em ab initio} numerical
calculations of the MAE to analyze the key physical quantities
determining the anisotropic magnetic characteristics of single
3$d$-metal atoms on CuN/Cu(001)-c(2x2) substrate. Similar to the
theory of  Ref. \cite{Hirjibehedin}, we use
a supercell model. The supercell consists of three Cu(001) layers
and a single Cu$_2$N atomic layer with c(2x2)N-Cu(001) arrangement
given in \cite{yoshimoto}. The in-plane c(2x2) dimentional unit cell
is doubled (Cu$_4$N$_2$), and the 3d-atom (Mn and Fe) is placed on
the top of Cu-atom .  The supercell is shown schematically in Fig.
1. The vacuum is modeled by the equivalent of four empty Cu layers.

The structure relaxation is performed employing the standard VASP
method \cite{vasp} without spin-orbit coupling (SOC) and making use
of the generalized gradient approximation. Placing 3d atom on the
top of the Cu atom in the CuN surface makes a substantial
rearrangement of the atomic structure (see Fig. 1). The Cu atom
right below the adatom moves toward the bulk and the relaxed
distance between this atom and magnetic atom is decreasing from 4.42
Bohr for the Mn atom to 4.27 Bohr for the Fe atom.
Other Cu atoms in the CuN top-layer change slightly their positions
with the change of the magnetic atom.
Overall relaxed atomic positions are qualitatively consistent with
the picture given in Ref. \cite{Hirjibehedin}

\begin{figure}[h]
\centerline{\includegraphics[angle=0,width=0.80\columnwidth,clip]{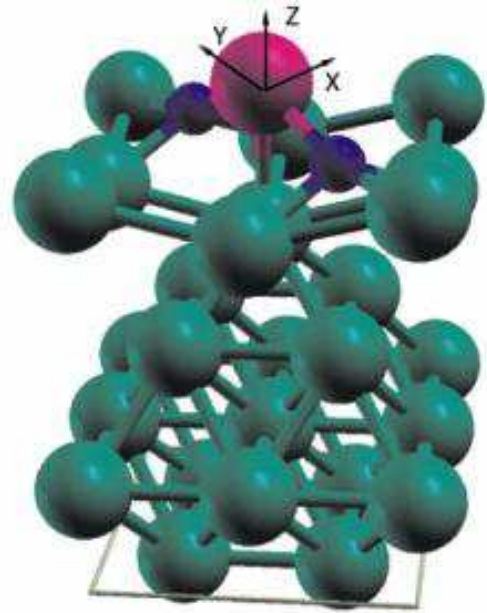}}
\caption{A schematic crystal structure used to represent the 3d-atom
on the c(2x2)N-Cu(001)surface. The actual atomic positions
correspond to the case of the Mn atom on the
c(2x2)N-Cu(001)surface.}
\end{figure}

\begin{figure}[t]
\centerline{\includegraphics[angle=0,width=1.0\columnwidth,clip]{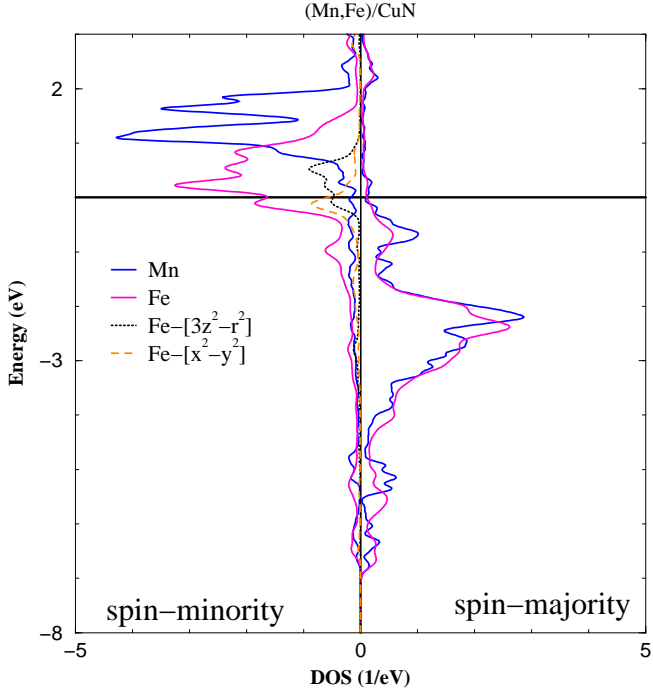}}
\caption{Spin-resolved PDOS for Mn and Fe adatoms. Also shown PDOS
for the Fe atom  ${x^2-y^2}$ and ${3z^2-r^2}$ spin-minority
orbitals.}
\end{figure}

We investigate the relativistic electronic and magnetic character of
3d-atoms on the c(2x2)N-Cu(001)surface. We use the relativistic
version of the full-potential linearized augmented plane-wave method
(FP-LAPW) \cite{singh}, in which spin-orbit (SO) coupling is
included in a self-consistent second-variational procedure
\cite{shick97}. The conventional (von Barth-Hedin) local
spin-density approximation is adopted in the calculations, which is
expected to be valid for itinerant metallic systems.

The spin $M_S$ and orbital $M_L$ magnetic moments for the
magnetization directed along the $z$-axis are given in Table~I. for
the Mn and Fe atoms. Small spin and orbital moments are also induced
on neighboring Cu sites and quickly decay away from the magnetic Mn
or Fe atom. The spin-resolved projected density of states (PDOS) for
the Mn and Fe atoms is shown in Fig.~2. The spin-majority manifold
is practically fully occupied for both Mn and Fe. For Mn atom, the
spin-minority channel is almost empty and the orbital $M_L$ moment
is almost zero. The spin-minority occupation is increased for the
Fe-adatom while the spin splitting and spin moment $M_S$ are
decreasing. The detailed inspection of $m_l$-projected PDOS shows
that non-zero orbital $M_L$ moment for the Fe atom originates from
$| m_s=-\frac{1}{2} ; m_l= \pm 2 \rangle$ orbitals near Fermi edge.
The major contribution to $M_L$ is brought about mainly by in-plane
$xy$ and ${x^2-y^2}$ spin-minority orbitals. The ${3z^2-r^2} (\sim
|m_l=0\rangle)$ spin-minority orbital (see Fig. 2) does not
contribute to $M_L$. This out-of-plane ${3z^2-r^2}$ orbital is the
least localized due to the strong overlap with 3$d$ electrons of the
Cu atom beneath.

%
\begin{table}[floatfix]
\caption{Total spin moment per unit cell (${M_S}^{Tot}$),  spin
($M_S$) and orbital ($M_L$) magnetic moments on 3d-adatom (in Bohr
magnetons) for the magnetization directed along the $z$-axis.}
\begin{center}
\begin{tabular}{ccccc}
\hline
\multicolumn{1}{c}{Atom}& ${M_S}^{Tot}$ & $M_S$ & $M_L$\\
Mn                      & 4.379            & 3.758 &0.004 \\
Fe                      & 3.654            & 2.917 &0.076 \\
\hline
\end{tabular}
\label{table1}
\end{center}
\end{table}

Next we turn to a sailent aspect of our investigation, the MAE
calculations. The anisotropic energy ${E_A}(\theta, \phi)$
dependence (including the second order terms) on the magnetization
direction  reads,
\begin{eqnarray}
\label{en} E_A(\theta, \phi) &=& K_2^{\perp} {\bf e}_z^{2} +
K_2^{||} ( {\bf e}_x^{2} - {\bf e}_y^{2}) \;,  \\ \nonumber
E_A(\theta,\phi) &=& K_2^{\perp} \cos^2(\theta) + K_2^{||}
\sin^2(\theta)(\cos^2(\phi) - \sin^2(\phi))
\end{eqnarray}
where $K_2^{\perp}$ and $K_2^{||}$ are the uniaxial MAE constants,
and $e_{x,y,z}$ are the cartesian coordinates of the normalized
magnetization vector $\vec{\bf M}/|\vec{\bf M}|$. The $\theta$ and
$\phi$ are the polar angles in the reference frame  which is chosen
as follows: the $x$-axis is along the in-plane hollow direction, the
$y$-axis is along the in-plane N-chain direction, and $z$-axis is
along the out-of-plane direction (see Fig. ~1).

In order to evaluate the MAE from Eq.(\ref{en}), we make use of the
torque method \cite{torque1} . It can be formulated as follows. We
solve the Kohn-Sham equations for a
two-component spinor $|\Phi_{i} \rangle = \left( \begin{array}{c} \Phi^{\uparrow}_i  \\
                                  \Phi^{\downarrow}_i \end{array} \right)$ \cite{sr},
\begin{eqnarray}
\sum_{\beta} \Big( -\nabla^{2} + \hat{V}_{eff} +  \xi ({\vec{\bf l}}
\cdot  {\vec{\bf s}} )  \Big)_{\alpha,\beta} \Phi^{\beta}_i({\bf r})
= e_{i} \Phi^{\alpha}_{i}({\bf r}) \; , \label{eq:2}
\end{eqnarray}
where the $\hat{V}_{eff} = V({\bf{r}})\mbox{$\hat{I}$} \;+\;
\mbox{\boldmath{$\sigma$}} \cdot {\bf B}({\bf{r}})$ matrix consists
of the sum of the scalar potential $V$  and ``exchange" field $B$
parallel to the spin moment $M_S$, and $\hat{H}_{\mbox{SO}} = \xi ({
\vec{\bf l}} \cdot {\vec{\bf s}})$ is the SO coupling operator. When
the magnetic force theorem \cite{force} is used to evaluate the
magnetocrystalline anisotropy energy, the $M_S$ is rotated and a
single energy band calculation is performed for the new orientation
of $M_S$. The MAE results from SO coupling induced changes in the
band eigenvalues $E_A(\theta,\phi) = \sum_{i}^{occ}
\epsilon_i(\theta,\phi)$. Alternatively, the torque $T(\theta,\phi)
= \partial E_A(\theta,\phi)/ \partial \theta$ can be evaluated
making use of the linear response theory:
\begin{eqnarray}
T(\theta,\phi) = \sum_{i}^{occ} \langle{\Phi'_i}|\frac{\partial {\bf
U}}{\partial \theta} \xi ({\vec{\bf l}} \cdot  {\vec{\bf s}}) {{\bf
U}^{\dagger}} + {\bf U} \xi ( {\vec{\bf l}} \cdot  { \vec{\bf s}})
\frac{\partial {\bf U}^{\dagger}}{\partial \theta} |\Phi'_i\rangle
\end{eqnarray}
where the ${\bf U}(\theta,\phi)$ is a conventional spin rotation
matrix and $|\Phi'\rangle = {\bf U}(\theta,\phi) |\Phi\rangle$. An
advantage of this approach is that it allows the total MAE
separation into the element-specific contributions from different
atoms in the unit cell. The torque method has been first implemented
in FP-LAPW basis in Ref. \cite{wang96}. Also, it has been employed
recently in the Korringa-Kohn-Rostocker calculations \cite{torque2}.

\begin{figure}[t]
\includegraphics[angle=0,width=1.0\columnwidth,clip]{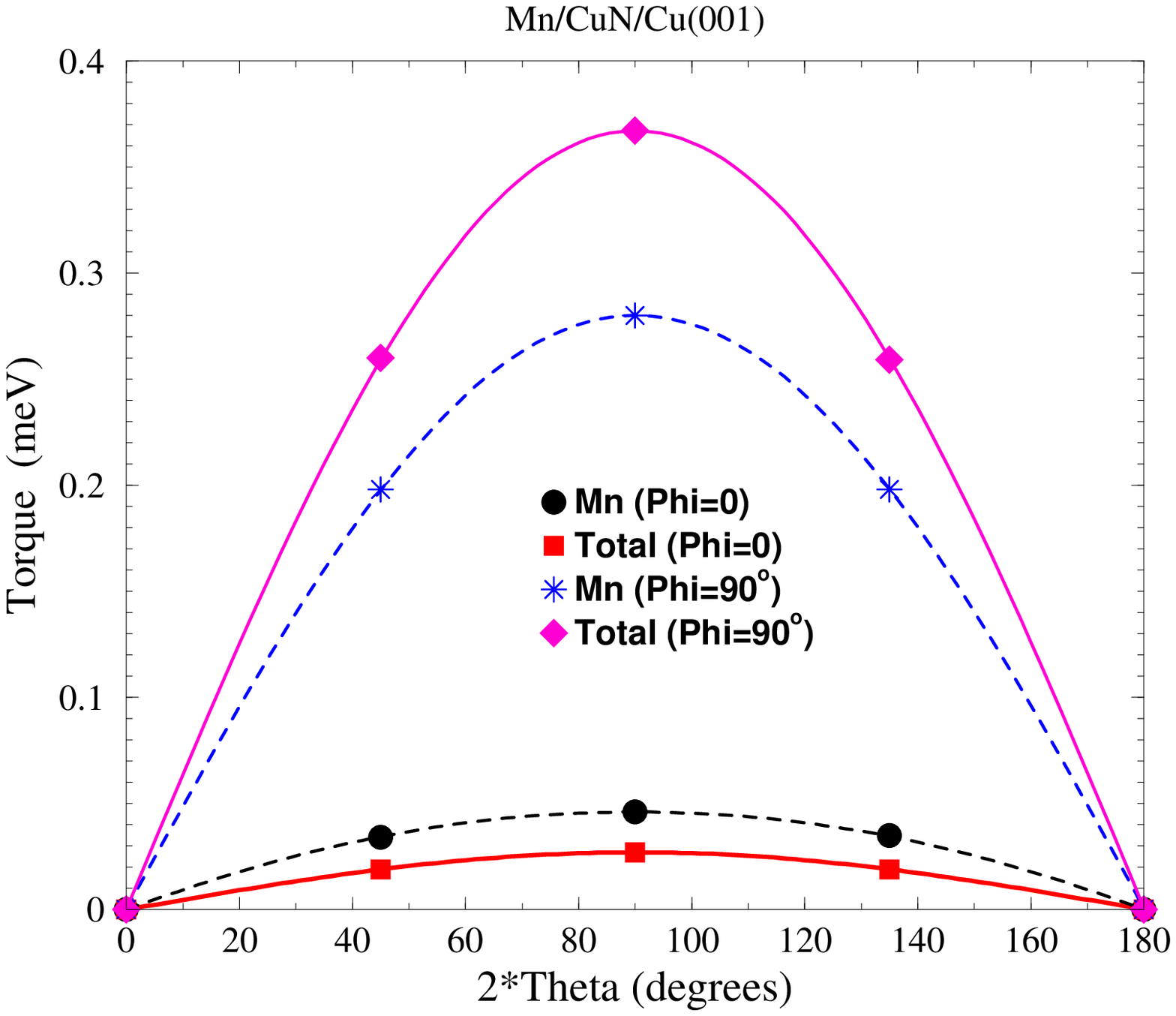}
\includegraphics[angle=0,width=1.0\columnwidth,clip]{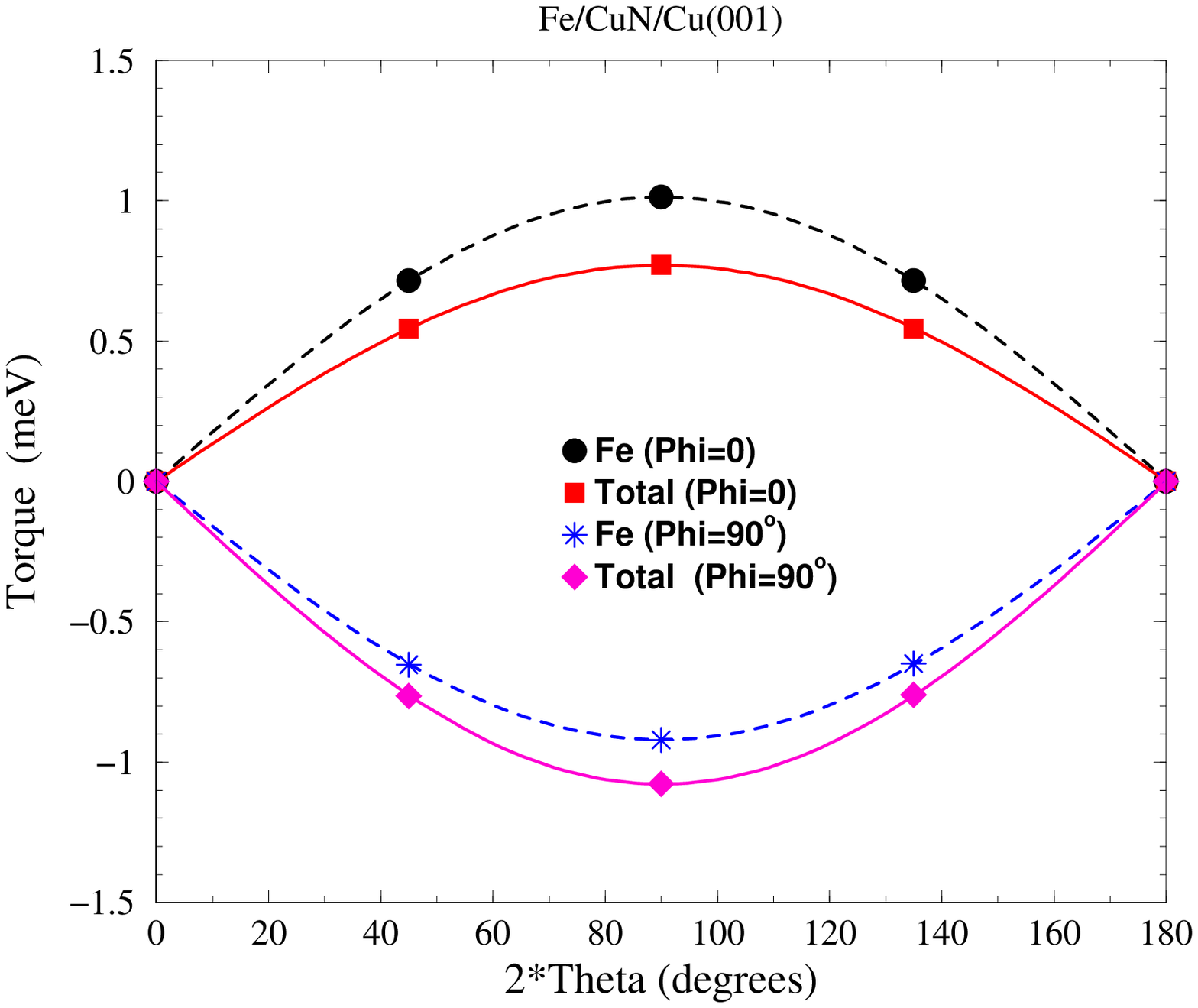}
\caption{The torque $T(\theta,\phi)$ for the $x-z$-plane
($\phi=0^o$) and $y-z$-plane ($\phi=90^o$) as a function of
$\theta$. The total torque and leading contributions from the
3d-adatom are shown.}
\end{figure}

 The torque $T(\theta,\phi)$ angular dependence is shown in Fig.~3 for both Mn and Fe-atoms on
CuN/Cu(001). A set of $784$ $k$-points in the full 2D-BZ which is
equivalent to  3136 $k$-points in the full 2D-BZ of Cu(001)) is used
in these calculations. The uniaxial MAE constants $K_2^{\perp}$ and
$K_2^{||}$ can be evaluated from the torque $T(\theta,\phi)$ angular
dependence, shown in Fig.~3 and angular derivative of Eq.(\ref{en}),
\begin{eqnarray}
\label{tq} T(\theta,\phi) &=& [-K_2^{\perp} + K_2^{||} \cos(2 \phi)]
\sin(2\theta) .
\end{eqnarray}
For the Mn atom, the values of the uniaxial MAE constants  are:
$K_2^{\perp}$=-0.20 meV and $K_2^{||}$=-0.17 meV. The Mn atom
contribution in $K_2^{\perp}$ = -0.16 meV, and $K_2^{||}$=-0.12 meV.
For the Fe atom, $K_2^{\perp}$=0.16 meV and $K_2^{||}$=0.93 meV, and
the Fe atom specific contribitions in $K_2^{\perp}$ = -0.04 meV, and
$K_2^{||}$=0.97 meV. Also, we found that higher order anisotropy is
much less (at least by an order of magnitude) than the uniaxial
anisotropy.

Now we evaluate the MAE defined as the energy $E_A(\theta, \phi)$
difference for different directions of the magnetization $M$. Using
the torque $T(\theta,\phi)$ angular dependence shown in Fig.~3, we
obtain MAE=$\int_{0}^{\pi/2} d\theta \; T(\theta,\phi)$. The values
of the MAE are shown in Table II. There is an increase of the MAE
from Mn to the Fe atom case. For the Mn atom, the easy magnetization
axis is directed along the surface normal $z$-axis, in agreement
with the experimental data \cite{Hirjibehedin}. For the case of Fe,
the easy magnetization is along the N-chain, also in agreement with
the experiment \cite{Hirjibehedin}. The anisotropic energy
$E_A(\theta, \phi)$ angular dependence for Mn and Fe atoms on CuN
surface together with the easy magnetization axis orientation is
illustrated in Fig. 4.

\begin{table}[h]
\begin{center}
\begin{tabular}{ccccc}
\hline
Total MAE &Mn & Fe    \\
$\Delta E_A[z-x]$ &-0.03 & -0.77 \\
$\Delta E_A[z-y]$ &-0.37 & 1.08 \\
$\Delta E_A[y-x]$ &0.34  &-1.86 \\
\hline
\end{tabular}
\label{table1}
\end{center}
\caption{The MAE (meV) for the Mn and Fe-adatoms, Here, $\Delta
E[i-j] = E_A[M||j] - E_A[M||i]$, $i(j)=x,y,z$}
\end{table}

\begin{figure}[floatfix]
\includegraphics[angle=0,width=7.8cm,clip]{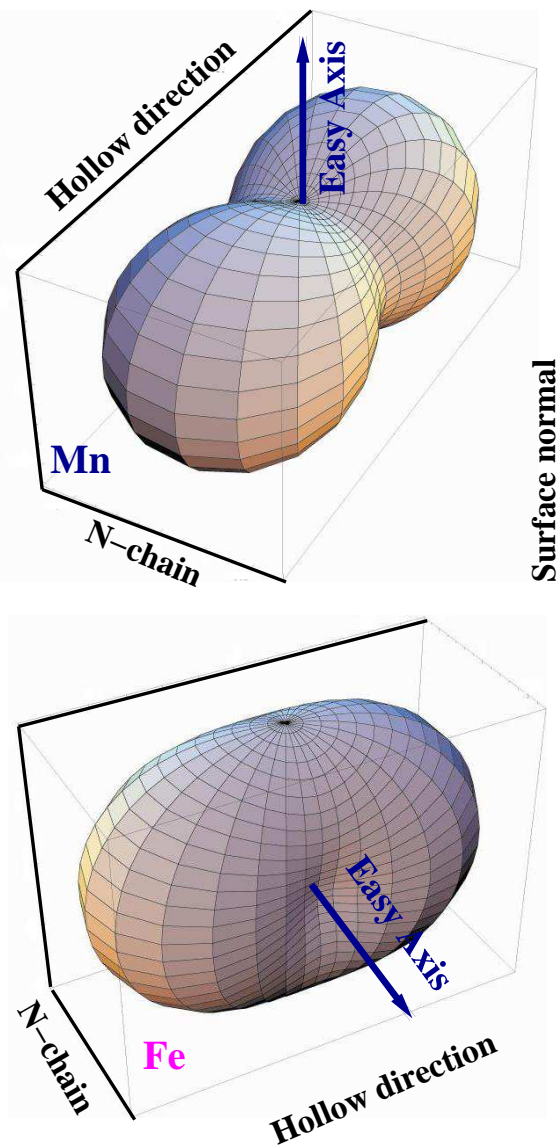}
\caption{The anisotropic energy \{$\theta,\phi$\} angular dependence
for Mn (top) and Fe (bottom) atoms on CuN surface. Here the length
of radius vector forming the surface is equal to $[E_A(\theta, \phi)
- E_A(\mbox{easy axis})]$.}
\end{figure}

It is quite common to examine the correlation between the MAE and
the orbital moment anisotropy  (OMA). Approximate relation between
the MAE and OMA is given by Bruno formula \cite{Bruno}, [MAE
$\approx -\xi/4$ OMA], where $\xi$ is the SOC constant. For the Mn
atom case, the Bruno formula gives $\Delta E_A[z-x]$ of -0.07 meV,
$\Delta E_A[z-y]$ of -0.02 meV, and $\Delta E_A[y-x]$ of -0.05 meV.
Comparison with the torgue results of Table~II. shows that Bruno
formula yields the correct easy z-axis but fails to describe the
${y-x}$ plane transverse anisotropy. For the case of Fe atom, making
use of Bruno formula we obtain the MAE of $\Delta E_A[z-x]$ = -0.74
meV, $\Delta E_A[z-y]$ = 1.46 meV, and $\Delta E_A[y-x]$ = -2.19 meV
in a good agreement with the torgue results (see Table~II.).
The reason why Bruno formula works better for Fe atom  than for Mn
atom case is that it is not accurate enough to account for
relatively small Mn atom MAE. For the stronger Fe atom MAE, validity
of Bruno formula is improving on qualitative.

Now we turn to comparison with the experimental results of
Hirjibehedin {\em et al.} \cite{Hirjibehedin}. The STM measures the
spin-excitation energies in a magnetic field. These excitation
spectra are then analysed by the model Hamiltonian,
\begin{eqnarray}
\label{h1} \hat{H} \; = g \mu_B {\bf B} {\bf S} + D {\bf S}_z^{2} +
E ( {\bf S}_x^{2} - {\bf S}_y^{2})
\end{eqnarray}
In the Eq.(\ref{h1}), the $z$-axis is chosen along the easy
magnetization direction. In order to compare our results with the
experiment, we have to convert the data in Table II. into the
reference frame chosen in Ref. \cite{Hirjibehedin} and re-normalize
the anisotropy values by $S^2$ ($S=5/2$ for Mn and $S=2$ for Fe).
The results are
shown in Table III. Our {\em ab initio} results correctly reproduce
the sign and order of magnitude of $D$ and $E$ experimental
anisotropies.

It is quite surprising that the LSDA based calculations give quite
reasonable values of the MAE constants for the systems which have
been initially thought as being close to the atomic limit. The
density functional theory is known to work reasonably well for the
ground state energy determination, and the MAE is defined as the
ground state energy difference for different magnetization
directions. Most probably, that is why the MAE results of density
functional theory resemble the values of the uniaxial MAE constants
experimentally determined from the spin excitation spectra of the
atomic spin. On the other hand, the electron correlation effects
beyond those which are already included in LSDA can play essential
role in more accurate theoretical modeling and interpretation of the
experimental data \cite{Hirjibehedin}. Further progress in realistic
calculations of the ground state properties and excitations for
single atomic spin on the surface will be made on the basis of the
newly emerging combination of the LSDA and dynamical mean field
theory \cite{Kotliar-RMP}.

In conclusion, we have shown that the magnetic anisotropy energies
for Mn- and Fe-atoms on CuN/Cu(001) can be semi-qualitatively
reproduced by the first-principles LSDA FP-LAPW calculations. The
easy magnetization direction is found in agreement with the
experimental data for Mn and Fe atoms. It is shown the calculated
MAE has a single-ion character and mainly originates from the well
localized moment of Mn- and Fe-atoms. The uniaxial MAE constants are
calculated in semi-quantitative agreement with the experiment.

We gratefully acknowledge discussions with K. von Bergmann, H.
Brune, and P.M. Oppeneer. Financial support was provided by the
Grant Agency of the Academy of Sciences (Project A100100530), DFG
Grant SFB668-A3 (Germany) and German-Czech collaboration program
(Project 436TSE113/53/0-1, GACR 202/07/J047).

\begin{table}[floatfix]
\caption{Comparison with experimental $D$ and $E$ (meV). Here, for
the Mn atom case, the $x$-axis is along N-chain, the $y$-axis is
along the hollow direction, and the $z$-axis is out-of-plane. For
the Fe atom case,  the $x$-axis is along the hollow direction , the
$y$-axis is out-of-plane, and the $z$-axis along N-chain
\cite{Hirjibehedin}.}
\begin{center}
\begin{tabular}{ccccc}
\hline
Mn & $D$ & $E$    & (meV)\\
Exp. &-0.039 $\pm$ 0.001 & 0.007 $\pm$ 0.001 \\
LSDA & -0.03 & 0.03 \\
\hline
Fe & $D$ & $E$    & (meV)\\
Exp. &-1.55 $\pm$ 0.01 & 0.31 $\pm$ 0.01 \\
LSDA &-0.36 &0.10 \\
\hline
\end{tabular}
\label{table1}
\end{center}
\vspace*{-0.5cm}
\end{table}



\vspace*{-0.25cm}
\begin{thebibliography}{99}
\vspace*{-1cm}
\bibitem{Hirjibehedin} S. Hirjibehedin {\em et al.},
Science 317, 1199 (2007).
\bibitem{gambardella03}
P. Gambardella {\em et al.}, Science 300 (2003) 1130.
\bibitem{yoshimoto} Y. Yoshimoto, S. Tsuneyuki, Surf. Sci. {\bf
514}, 200 (2002).
\bibitem{vasp} G. Kresse and J. Hafner, Phys. Rev. B 47, R558 (1993);
G. Kresse and J. Furthmüller, Comput. Mater. Sci. 6, 15 (1996); G.
Kresse and D. Joubert, Phys. Rev. B 59, 1758 (1999).
\bibitem{singh} D.J. Singh, {\em Planewaves, Pseudopotentials and the
LAPW Method} (Kluwer Academic, Boston, 1994), p. 115.
\bibitem{shick97} A.B. Shick, D.L. Novikov, and A.J. Freeman, Phys. Rev. B {\bf 56},
R14 259 (1997).
\bibitem{torque1} S. A. Turzhevskii, A. I. Lichtenstein and M. I. Katsnelson,
Soviet Physics - Solid State,  {\bf 32}, 1138 (1990).
\bibitem{sr} To simplify the notation, we use Pauli-like
Hamiltonian including SOC, while the actual implementation contains
in addition the scalar-relativistic terms.
\bibitem{force} A. I. Lichtenstein,
M. I. Katsnelson, V. P. Antropov, and V. A. Gubanov, J. Magn. Magn.
Mater. {\bf 67}, 65 (1987); M. Methfessel, and J. Kubler, J. Phys. F
{\bf 12}, 141 (1982).
\bibitem{wang96} X. Wang, R. Wu, D.-S. Wang, and A. J. Freeman,
Phys. Rev. {\bf B 54}, 61 (1996).
\bibitem{torque2} See for review J. Staunton {\em et. al.}, $\Psi_k$ -
network highlight (2007).
\bibitem{Bruno} P. Bruno, Phys. Rev. {\bf B 39}, 865 (2004).

\bibitem{Kotliar-RMP} A. Georges, G. Kotliar, W. Krauth, M. Rozenberg, Rev. Mod. Phys. {\bf 68}, 13
(1996).




\end{thebibliography}
\end{document}